\documentclass[twocolumn,showpacs,amssymb,prd]{revtex4}

\usepackage{graphicx}
\usepackage{dcolumn}
\usepackage{bm}
\usepackage{epsfig}

\begin{document}
\newcommand{\gsim}{\hbox{\rlap{$^>$}$_\sim$}}


\title{Origin Of The High Energy Cosmic Neutrino Background}

\author{Shlomo Dado} \affiliation{Department of Physics and Space Research 
Institute, Technion, Haifa 32000, Israel}

\author{Arnon Dar} \affiliation{Department of Physics and Space Research 
Institute, Technion, Haifa 32000, Israel}

\begin{abstract}

The diffuse background of very high energy extra-terrestrial neutrinos 
recently discovered with IceCube is compatible with that expected from 
cosmic ray interactions in the Galactic interstellar medium plus that 
expected from hadronic interactions near source and in the intergalactic 
medium of the cosmic rays which have been accelerated by the jets  that 
produce gamma ray bursts (GRBs).

\end{abstract}


\maketitle

\section{Introduction}

Based on combined 3-year data, the IceCube collaboration has reported 
recently$^1$ the discovery of a diffuse background of high energy (HE) 
astrophysical neutrinos with energy above 60 TeV that extends at least up 
to 3 PeV, in excess (5.7$\sigma$ significance level) of that produced by 
cosmic ray (CR) interactions in the atmosphere. The neutrino events did 
not point back to any identified sources and are consistent with an 
isotropic distribution of arrival directions. Their best fit power-law to 
the energy flux per flavor above 60 TeV is
\begin{equation} 
E^2\,\phi_\nu = 1.5\times 10^{-8} \left({E\over 100\,{\rm 
TeV}}\right)^{-0.3} {{\rm GeV\,\over cm^2\, s\, sr}}\, . 
\end {equation}

So far, the origin of this diffuse background of high energy neutrinos is 
not known. A natural source of high energy astrophysical neutrinos is the 
decay of mesons produced in collisions of CR nuclei with matter and 
radiation. Several models of cosmic ray production of neutrinos in/near 
cosmic ray sources$^2$ have been used to estimate the astrophysical 
diffuse background of high energy neutrinos. But, uncertain assumptions 
and free adjustable parameters prevented meaningful conclusions. Moreover, 
no evidence of neutrino emission from point-like or extended CR sources 
was found in four years of IceCube data$^3$. These include supernova 
remnants, the nearest most luminous gamma ray bursts (GRBs) such as$^4$ 
130427A at redshift $z=0.34$, the giant elliptical galaxy M87 at a 
distance of 16 Mpc that contains the nearest active galactic nucleus (AGN) 
in the northern sky, and Markarian 421, at a distance of 133 Mpc, the 
brightest $\gamma$-ray blazar in the northern sky. In particular, the null 
detection of high energy neutrinos with IceCube from 100 stacked GRBs has 
been used to set a strong limit on the contribution of GRBs to the diffuse 
background of high energy neutrinos$^5$.  But, this flux limit was based 
on the assumption that the neutrino emission in GRBs roughly coincides in 
time and beaming with the prompt gamma-ray and early afterglow emissions, 
as expected$^6$ in the fireball models of GRBs.

GRB neutrinos, however, may be produced mainly through {\it hadronic 
production of mesons by GRB cosmic rays in the giant molecular clouds 
(GMCs) where most core colapse SNe take place} rather than through 
photo-production of mesons, which are assumed to take place in GRB 
"fireballs" during their prompt gamma ray and early afterglow 
emissions$^6$. Such hadronic production of mesons, which is expected in 
the cannonball model of GRBs$^7$, is delayed and spread over much larger 
beaming angles than the prompt gamma ray and early afterglow emissions 
(the early time GeV emission from GRBs observed with Fermi-LAT, most 
probably, is not from $\pi^0$ decay but from inverse Compton scattering by 
the high energy electrons of their self emitted synchrotron 
radiation$^8$). A large spread in the neutrino arrival times (due to CR 
diffusion through the GMC) reduces the detected flux from individual GRBs 
and stacked GRBs. Hence, limits on the flux of GRB neutrinos$^5$ based on 
no detection in selected time and sky windows centered on visible GRBs may 
not be applicable to the diffuse flux of neutrinos from all past GRBs, 
which may contribute significantly and even dominate the isotropic 
component of the diffuse background of high energy neutrinos.

Moreover, in this letter, we show that the spectrum, sky distribution and 
intensity of the diffuse background of high energy neutrinos recently 
discovered with the IceCube detector$^1$ are compatible with those of a 
non-isotropic Galactic background expected from hadronic collisions of 
very high energy cosmic rays in the Galactic interstellar medium 
(ISM) plus an isotropic extragalactic background expected from hadronic 
collisions near source and in the intergalactic medium (IGM) of high 
energy cosmic rays accelerated by the highly relativistic jets, which 
produce GRBs in star forming galaxies.

\section{The Galactic Background}

Because of Feynman scaling$^9$, very high energy (VHE) cosmic rays with a 
power-law flux $\phi_p\propto E^{-k}$ per-nucleon that collide with the 
baryons of an "optically thin" gas produce mesons whose leptonic and 
semileptonic decays produce neutrinos at a rate $\dot{n}_\nu\approx 4\,\pi 
K\,\bar{\sigma}_{in}(pp)\,\phi_p\, n_b$ per unit volume.  The constant $K$ 
depends$^{10}$ only on the power-law index $k$, $\sigma_{in}(pp)$ is the 
inelastic pp cross section, and $n_b$ is the baryon density of the gas. 
Since the average energy of the produced neutrinos is $\sim\! 1/50$ that 
of the CR nucleons, production of neutrinos in the 60 TeV - 3 PeV 
energy-range requires CR nuclei with energy {\it per nucleon} well above 
the CR "knee"  where $\sigma_{in}(pp)\approx 62\,E_{PeV}^{0.085}$ mb. The 
CR flux {\it per nucleon} in the 100 PeV - 10 EeV energy range is 
roughly$^{11}$ $\phi_p=0.07\, E_{GeV}^{-2.67}\,{\rm 
GeV^{-1}\,cm^{-2}\,sr^{-1}\,s^{-1}}$ with $E_{GeV}=E/$GeV. This flux 
represents quite well the CR flux {\it per nucleon} between the CR knee 
and CR ankle. Presumably, these very high energy (VHE) CRs are 
accelerated by the highly relativistic jets of supernovae of type Ic that 
produce GRBs, most of which point away from Earth$^{12,13}$.

Most of the gas and dust in the Galaxy ($M_{gas}\approx 5.5\times 10^9\, 
M_\odot$) resides in a relatively thin Galactic disc whose mean surface 
density within $\sim\! 12.5$ kpc from its center is$^{14}$ 
$\Sigma_{gas}\!\approx\!10\, M_\odot\, {\rm pc^{-2}}$,  and beyond it the 
mid plane density drops exponentially in the radial direction like $n_0\, 
exp[-(R-R_\odot)/R_s]$ with $n_0=0.9\, {\rm cm^{-3}}$ with a scale height 
$R_s=3.75\,$ kpc and where $R_\odot\sim 8.5$ kpc is the distance of 
the solar system 
from the center of the disc. This gas and dust are embedded in a large 
cosmic ray halo with roughly a constant density in the Galactic disc. 
Hence, the total neutrino luminosity of the Galaxy is given roughly by 
\begin{equation} 
dL_\nu \approx K\, \bar{\sigma}_{in}(pp)\, 
(M_{gas}/m_p)\,4\,\pi\,\phi_p(E)\,E\,dE 
\end{equation} 
where $K=0.06$ for 
$k=2.7$. A rough estimate of the neutrino event rate in IceCube due to 
Galactic $\nu's$ with $E_\nu\!>\!60$ TeV, is 
\begin{equation} 
\dot{N}_\nu\approx K\, \int d^3{\bf R}\,{n_b({\bf R}) \over {\bf 
|R-R_\odot|^2}}\, \int A\, \sigma_{in}\,4\,\pi\,\phi_p\, dE\, 
\end{equation} 
where $A(E)$ is the mean effective area of IceCube for 
neutrinos of the $e, \mu,\tau$ flavors (assuming equal mixing due to 
oscillations). For an energy-weighted$^2$ $<A(E)/E>=24.5\, {\rm 
cm^{2}/GeV}$ above 60 TeV, and $n_b({\bf R})$ as parametrized in reference 
14, Eq.(3) yields $\dot{N}_\nu\approx 3.6\,{\rm y^{-1}}$ neutrino events 
pointing back to the Galactic disc and peaked towards the Galactic center. 
The contribution of CRs accelerated in episodes of mass accretion onto the 
Galactic central massive black hole may contribute significantly to this 
concentration. All together, the energy spectrum of these Galactic 
neutrino background is expected to have an $\sim 
E^{-2.67+0.085}\approx E^{-2.58}$ spectrum, and 
its sky distribution is expected to follow roughly that of the diffuse 
Galactic gamma rays in the energy range 100 MeV - 100 GeV, which are 
produced mostly by hadronic $\pi^0$ production and leptonic bremsstrahlung 
in the Galactic disc and are not absorbed significantly by the background 
light in the Galaxy.

\section{GRB Neutrinos} 

There is mounting observational evidence that long duration GRBs and their 
afterglows are produced by the interaction of highly relativistic jets of 
plasmoids (cannonballs) of ordinary matter ejected in stripped envelope 
supernova explosions, mainly of type Ic, with the radiation and matter 
along the jets' trajectories$^{13}$. In particular, the prompt 
$\gamma$-ray pulses and early-time X-ray flares are produced by the jet 
electrons through inverse Compton scattering (ICS) of photons in a light 
halo (glory)  surrounding the progenitor star. Such a light halo is formed 
by scattered stellar light from ejecta of pre-supernova eruptions$^{15}$. 
Thus, the total kinetic energy of the baryons in the jet is $E_k\!\sim\! 
(m_p/m_e)$ times $E_\gamma$, the total emitted gamma ray energy. The jet 
decelerates by the swept in particles in front of it. Its random magnetic 
fields transform its kinetic energy by Fermi acceleration to CRs' energy 
with a spectrum $E^{-2}$. The CRs escaoe the jet by diffusion through its 
random magnetic fields. For a Kolmogorov spectrum$^{25}$ of the random 
magnetic fields, their diffusion coefficient is $\propto\! E^{1/3}$, which 
yields a residence time $\tau(E)\!\propto\!E^{-1/3}$ and a CR flux  
$A\, E_{GeV}^{-2.33}$ where $A\approx E_k/3\, GeV $.

The CR jets move through the GMCs with practically the speed of light and 
produce narrow beams of high energy neutrinos mainly through $\pi$ and $K$ 
production in hadronic collisions with the baryons along their path in the 
GMCs. Under the assumption of Feynman scaling$^9$, CR spectrum 
$\sim\!E^{-2.33}$, and a baryonic column density $N_p$ of the GMC, 
roughly, a fraction$^{16}$ $f_\nu\approx 0.09\,\sigma_{in}(pp)\,N_p$ of 
the CR flux is converted to $\nu_\mu+\nu_e$ flux with the same $\sim\! 
E^{-2.33}$ power-law energy spectrum.

Star formation in GMCs in the local universe (redshift $z=0$) seems to 
have a threshold around a surface density$^{17}$ $\Sigma_{th}\approx (129 
\pm 14)\, M_\odot\,pc^{-2}$, i.e., a baryon column density $N_p\approx 
1.63\times 10^{22}\, {\rm cm^{-2}}$. Such a column density is consistent 
with the mean column density $N_p=(8\pm 2)\times 10^{21}\,(1+z)^{1.25}{\rm 
cm^{-2}}$ inferred$^{18}$ from X-ray absorption after correcting it for 
the observed metallicity$^{19}$ of GRB host galaxies 
$Z(z)=0.54\,(1+z)^{-1.25}\, Z_\odot$, i.e., $N_p(z)\approx 1.47\times 
10^{22}\, (1+z)^{2.5}{\rm cm^{-2}}$. Hence, for a ballistic motion of the 
jet with approximately the speed of light through the GMC, we adop 
$f_\nu\approx 7.8\times 10^{-5}\, (1+z)^{2.5}\,(E/100{\rm TeV})^{-0.085}$.

Assuming isotropic emission, the gamma ray energy emitted by a GRB is 
$Eiso=4\, \pi\, D_L^2(z)\,F_\gamma/(1+z)$ where $D_L(z)$ is the GRB 
luminosity distance in a flat $\Lambda$CDM universe with a Hubble constant 
$H_0=67.3$ km/s Mpc, matter density $\Omega_M=0.315$, dark energy density 
$\Omega_\Lambda=0.685$ and baryon density $\Omega_b=0.049$, all in 
critical mass units$^{20}$, and $F_\gamma$ is the observed $\gamma$-ray 
fluence from the GRB. If the $\gamma$-ray energy is beamed into a solid 
angle $\Delta \Omega$, then the true GRB gamma-ray energy is $Eiso\, 
\Delta\Omega/4\, \pi$, and the true GRB rate is $\dot{N}_{GRB}\, 4\, \pi / 
\Delta\Omega $. Consequently, the total power in $\gamma$-ray emission by 
GRBs at redshift $z$ is
\begin{equation}
Eiso\, {\Delta\Omega\over 4\, \pi}\, d\dot 
{N}_{GRB}(z)\, {4\, \pi \over \Delta\Omega}= Eiso\,d\dot{N}_{GRB}(z), 
\end{equation} 
{\it independent of the beaming angle of GRBs}.

Noting that the full sky rate of GRBs is$^{21}$ $\dot{N}_{GRB}\!\approx\! 
3/day$, the flux of GRB neutrinos of all flavors per sr at Earth can be 
written as 
\begin{equation} 
E^2\phi_\nu \approx {\dot{N}_{GRB}\over N_{GRB}}{m_p\over m_e}
\Sigma_i {f_\nu(z_i)\,(1\!+\!z_i)^{2/3} Eiso(z_i)
\over 12\, \pi\, [D_L(z_i)]^2}\,\left[{E\over 
m_p}\right]^{-1/3}
\end{equation} 
where the summation 
extends over all $N_{GRB}=136$ long duration GRBs with known redshift 
and $Eiso$, which were reported before May 15, 2014  in the 
Greiner Catalog of GRBs$^{22}$ and in the GCN 
Circulars Archive$^{23}$, respectively. The distribution of such long GRBs 
as function of $z$ is shown in Fig.~1. 
\begin{figure}[] 
\centering
\epsfig{file=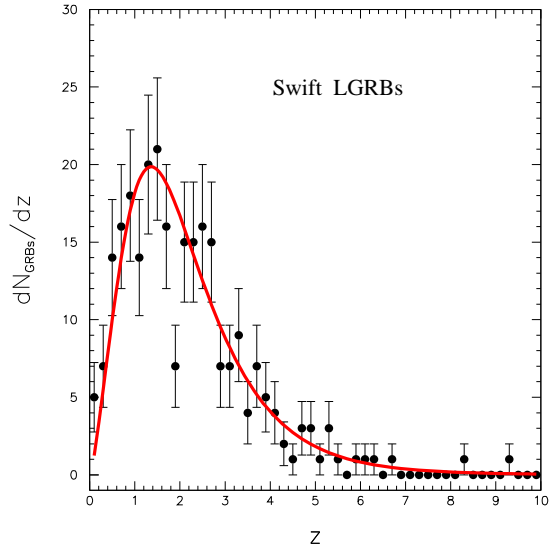,width=8.cm,height=8.cm} 
\caption{The redshift 
distribution of the 262 Swift LGRBs with known redshift that were detected 
before November 15, 2013 and their expected distribution$^{24}$ assuming 
the rate of GRBs/SNe traces the star formation rate.} 
\label{Fig1} 
\end{figure} 
For that distribution, Eq.~(5) yields  $E^2\phi_\nu\!\approx\! 
0.58 \times10^{-8}\,(E/100\,{\rm TeV})^{-0.25}\, {\rm GeV/cm^2\, s\, sr}$ 
{\it per neutrino flavor}.  

To a very good approximation, all the VHE cosmic ray nuclei from GRBs 
reach shortly (on a cosmic time-scale) the
IGM without any energy loss. In the IGM,  they continue to produce 
neutrinos in collisions with the IGM 
baryons. Most of the baryons in the universe ($\!>\! 90\%)$ reside 
in the IGM with a density $n_b(z)\!=\!2.5\times 10^{-7}\, 
(1+z)^3\, {\rm 
cm^{-3}}$. Hence, the effective column density encountered by a CR  
nucleon, which  is ejected into the IGM at redshift $z$, is 
$Np(z)\!\approx\!n_b(0)\,(c/H_0)\,\int_0^z dz'(1+z')^3/ (1+z')
\sqrt{(1+z')^3\Omega_M\!+\!\Omega_\Lambda}$. Using a conversion 
coefficient 
$f_\nu(z_i)$ for such $N_p(z)$ in Eq.~(5), the IGM contribution to the 
diffuse VHE energy isotropic neutrino background due to GRBs is  
$E^2\phi_\nu\!\approx\!
0.11 \times10^{-8}\,(E/100\,{\rm TeV})^{-0.25}\, {\rm GeV/cm^2\, s\, 
sr}\,\,$ {\it per neutrino flavor}. 

\section{Extragalactic ISM Neutrinos}
 
The bolometric luminosity per unit volume in the local universe  
($z\!=\!0$) is$^{26}$ $LD(0)\!\approx\! 1.4\times 10^8\,L_\odot/Mpc^3$. 
The bolometric luminosity 
of our Milky Way (MW) galaxy is $L_B(MW)\approx 2.3\times 
10^{10}\,L_\odot$. The ratio $LD(0)/L_B(MW)\approx 6.0\times 10^{-3}\, 
{\rm Mpc^{-3}}$ is approximately also the observed ratio of their core 
collapse supernova rates and consequently also of their GRB rates. 
Assuming that the VHE neutrino luminosity of external galaxies (EG) due 
to hadronic CR interactions in their ISM is proportional to their 
GRB rates, then a rough estimate of the cummulative neutrino flux from 
all external galaxies is  
\begin{equation} 
E^2\phi_\nu(EG)\approx {\,4\pi\,  LD(0)\,R_{MW}^2\, 
c\over  3\, L_B(MW)\, H_0}\,2.9\, E^2\phi_\nu(MW) 
\end{equation} 
The factor 2.9 is the effective value of $N_p(z)/N_p(0)\approx 
(1+z)^{2.5}$ inferred from GRBs X-ray absorption in the GRB host galaxies, 
weighted by the  star formation history$^{27}$ as in Eq.~(5). 
Eq.~(6) yields an estimated 
$E^2\phi_\nu(EG)\approx 0.03\times 
\,E^2\bar{\phi}_\nu(MW)$  contribution to the diffuse extragalactic 
background 
of  VHE neutrinos, where $E^2\bar{\phi}_\nu(MW)$ is the average Galactic 
neutrino background per sr.\\ 

\section{conclusions} 
Meson production in hadronic interactions of very 
high energy cosmic rays in the Galactic ISM is expected to produce a 
diffuse Galactic background of very high energy neutrinos with an energy 
flux $E^2\phi_\nu\!\propto\! E^{-2.58}$. The 
predicted number of Galactic neutrino events (all flavors)  in IceCube 
with $E\!>\!60$ TeV is roughly 3.6 events per year.
Their neutrino arrival directions 
are expected to trace the ISM column density in their arrival directions, 
i.e., point back mainly to the Galactic disc with a concentration 
towards the Galactic center. Their sky distribution is expected to follow 
closely that of the diffuse Galactic gamma ray background in the energy 
range 100 MeV - 100 GeV where the ISM is transparent to the $\gamma$-rays 
that are produced mostly by $\pi^0$ decay and leptonic bremsstrahlung. 

Meson production in hadronic interactions of very high energy cosmic rays 
accelerated by the highly relativisic jets, which produce GRBs within 
giant molecular clouds in star forming galaxies, where most GRBs take 
place, is expected to produce an isotropic background of very high energy 
neutrinos with a per flavor  flux $E^2\phi_\nu\!\approx\! 0.64\times 
10^{-8} (E/100\,{\rm TeV})^{-0.25}\, {\rm GeV/ cm^2\, s\,sr}$. Such an 
energy flux produces $\sim\! 3$ neutrino events per year with $E\!>\!60$ 
TeV in the IceCube detector.

Cosmic ray interactions in the ISM of external galaxies are expected to 
contribute to the {\it isotropic} background of very high energy neutrinos 
a per flavor energy flux $E^2\phi_\nu\!\approx\! 0.03\times 10^{-8} 
(E/100\,{\rm TeV})^{-0.58}\, {\rm GeV/ cm^2\, s\,sr}$. Such an energy flux 
produces $\sim\! 0.1$ neutrino events per year with $E\!>\!60$ TeV in the 
IceCube detector.

Fig.~2, compares the 
per-flavor flux$^1$ of VHE neutrinos with energy above 60 TeV
measured with IceCube$^1$
and the predicted flux due 
to the hadronic interaction of cosmic rays in the Galactic ISM 
(represented 
there by an equivalent isotropic MW flux), in giant molecular 
clouds of star forming galaxies where most core collapse supernovae, and 
hence GRBs, take place, and in the IGM. 
It shows that the predicted flux is 
compatible with that observed by IceCube. 
Special effort was made to base our
estimates  only on general considerations and 
priors, and avoid completely free adjustable parameters. 
However, in view of the
large uncertainties in the values of the priors, the good agreement 
between the expected  and the observed  neutrino flux by IceCube
cannot be used to draw firm conclusions beyond the statement that they are 
compatible. Obviously, much larger statistics are 
needed to test conclusively whether the diffuse background of VHE 
neutrinos is a sum of a Galactic non-isotropic background with a 
spectrum $\sim E^{-2.58}$ and intensity proportional to the Galactic 
column density in their arrival direction,
plus an isotropic extragalctic background with a harder 
spectrum $\sim E^{-2.25}$. 
\begin{figure}[] 
\centering
\epsfig{file=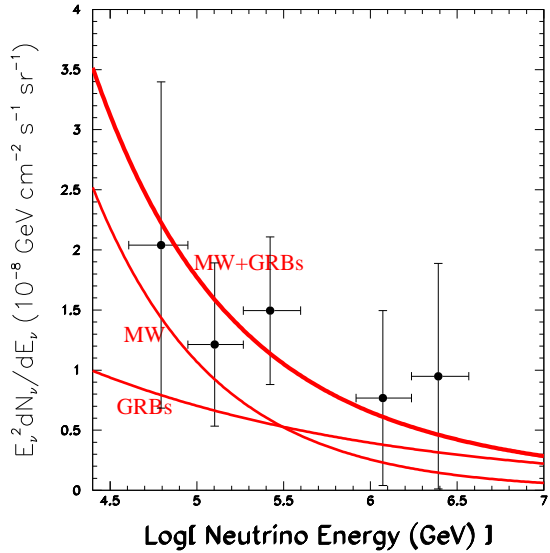,width=8.cm,height=8.cm} 
\caption{Comparison 
between the flux of an isotropic neutrino background that was inferred 
from 3 years measurements with IceCube$^1$, and that expected from cosmic 
ray production in the Galactic ISM and in the molecular clouds of external 
galaxies hosting supernovae that produce GRBs.} 
\label{Fig2} 
\end{figure}

{\bf Acknowledgment:} The authors thank E. Behar, L. Green, D. Guetta and 
S. Nussinov for useful remarks and an anonymous referee for very useful 
comments and suggestions.


\begin{thebibliography}{999}

\bibitem{Aarsten}%
M. G. Aartsen, et al. (The IceCube Collaboration), 
e-print arXiv:1405.5303. M. G. Aartsen, et al. (The IceCube 
Collaboration), {\it Science} {\bf 342}, 1242856 (2013).

\bibitem{Anchordoqui}%
For a recent review see L.  Anchordoqui, et al. 
{\it J. High Ener. Phys.} {\bf 1-2}, 1 (2014). See also,
Ruo-Yu Liu et al., {\it Phys. Rev. D} {\bf 89}, 083004 (2014).

\bibitem{Abassi1}%
R. Abbasi, et al. (IceCube Collaboration), {\it Astrophys. J.} 
{\bf 732}, 18 (2011); M. G. Aartsen, et al. (IceCube 
Collaboration), {\it Astrophys. J.} {\bf 779}, 132 (2013); M. G. Aartsen, 
et al. (The IceCube Collaboration), e-print arXiv:1406.6757.

\bibitem{Blaufuss,}%
E. Blaufuss, et al. (IceCube Collaboration) GCN Circ. 
14520

\bibitem{Abassi2}%
R. Abbasi, et al. (IceCube Collaboration), {\it 
Nature,} {\bf 484}, 351 (2012).

\bibitem{Waxman03}%
E. Waxman and J. Bahcall, {\it Phys. Rev. Lett.} {\bf 
78}, 2292 (1997); J. P. Rachen and P. Meszaros, {\it Phys. Rev. 
D}, {\bf 
63}, 023003 (2000); C. D. Dermer,{\it Astrophys. J.} {\bf 574}, 65 (2002); 
D. Guetta, et al. {\it Astroparticle Phys.} {\bf 20}, 409 (2004); L. A. 
Anchordoqui, et al. {\it Astroparticle Phys.} {\bf 29}, 1 (2008), I. 
Cholis, and D. Hooper, {J. Cos. Astroparticle Phys.} {\bf 06}, 030 (2013); 
K. Murase and K. Ioka, {\it Phys. Rev. Lett.} {\bf 111}  121102 (2013);
and references therein.

\bibitem{Dar01}%
See, e.g., A. Dar and A. De R\'ujula, {\it Phys. Rept.} 
{\bf 466}, 179 (2008) and references therein.

\bibitem{Dado01}%
S. Dado and A. Dar, e-print arXiv:0910.0687

\bibitem{Fynman}%
R. P. Feynman, {\it Phys. Rev. Lett.} {\bf 23}, 
1415-1418 (1969).

\bibitem{Dar04}%
A. Dar, {\it Phys. Rev. Let.} {\bf 51}, 227 (1983).


\bibitem{Apel}%
W. D. Apel et al. (KASCADE-Grande Collaboration) {\it 
Phys. Rev. D} {\bf 87}, 081101 (2013); I. C. Maris (Pierre Auger 
Collaboration) {\it EPJ Web of Conferences} {\bf 53}, 04002 (2013).


\bibitem{Dar02}%
A. Dar and R. Plaga, {\it Astron. \& Astrophys.} {\bf 
349}, 259 (1999).


\bibitem{Dar03}%
A. Dar and A. De R\'ujula, {\it Phys. Rept.} {\bf 405}, 
203 (2004); S. Dado, A. Dar and A. De R\'ujula, A.  {\it Astrophys. J.} 
{\bf 693}, 311 (2009), and references therein.

\bibitem{Kalberla}%
P. M. Kalberla and L. Dedes, {\it Astron. \& 
Astrophys.} {\bf 487}, 951 (2008).


\bibitem{Graham}%
See, e.g., M. L. Graham, et al. e-print arXiv:1402.1765



\bibitem{Bergin}%
Bergin, E. A. \& Tafalla, M. {\it Ann. Rev. Astron. 
Astrophys.} {\bf 45}, 339 (2007); A. Heiderman et al. {\it Astrophys. J.} 
{\bf 723}, 1019 (2010).



\bibitem{Dar05}%
A. Dar, {\it Phys. Let. B}, {\bf 159}, 205 129 (1985).





\bibitem{Campana}%
S. Campana et al.  et al. {\it Mont. Not. Roy. Astrn. 
Soc.} {\bf 402}, 2429 (2010); P. A. Evans, et al. {\it. Mon. Not. Roy. 
Ast. Soc.} {\bf 397}, 1177 (2009).

\bibitem{Savaglio}%
S. Savaglio, K. Glazebrook and D. LeBorgne, {\it 
Astrophys. J.} {\bf 691}, 182 (2009); C. C. Thone, et al. {\it 
Astronomische Nachrichten} {\bf 332}, 281 (2011), and references therein.

\bibitem{Ade}%
P. A. R. Ade, et al. Planck Collaboration:  e-print, 
arXiv:1303.5076


\bibitem{Stern}%
B. E. Stern, et al.  {\it Asrophys. J.}, {\bf 563}, 80 
(2001).


\bibitem{Greiner}%
J. Greiner, http://www.mpe.mpg.de/~jcg/grbgen.html


\bibitem{Barthelmy}%
S. Barthelmy, http://gcn.gsfc.nasa.gov/gcn-main.html

\bibitem{Dado2}%
S. Dado and A. Dar, {\it Asrophys. J.}, {\bf 785}, 70 
(2014).


\bibitem{Kolmogorov}%
A. Kolmogorov, {\it Dokl. Akad. Nauk SSSR}, {\bf 
30},301 (1941) [reprinted in {\it Proc. R. Soc. London} {\bf A 434}, 9 
(1941)].


\bibitem{Vaccari}%
M. Vaccari, et al. {\it Astron. \& Astrophys.} {\bf 
518}, L20 (2010).



\bibitem{Hopkins}%
We have used the parametrization in ref.~25  of the star 
formation rate compiled by A. M. Hopkins and J. F. Beacom, {\it 
Astrophys. J.} {\bf 651}, 142 (2006); N. A. Reddy and C. C.  Steidel, {\it Astrophys. 
J.} {\bf 692}, 778, (2009).

\end{thebibliography}
\end{document}